\documentclass[11pt,twoside]{article}
\usepackage[pdftex]{graphicx}
\usepackage{amsmath}
\usepackage{amssymb}
\usepackage{cite}

\newcommand{\w}{\mathit w}
\DeclareMathOperator{\tr}{tr}

\begin{document}
\newcommand{\pst}{\hspace*{1.5em}}

\newcommand{\rigmark}{\em Journal of Russian Laser Research}
\newcommand{\lemark}{\em Volume 30, Number 5, 2009}

%\lhead[\fancyplain{\rigmark, {\em \lemark}}{\rigmark}]{\fancyplain{\rigmark, {\em \lemark}}{\lemark}}
%\chead{}\rhead[\fancyplain{}{\lemark}]{\fancyplain{}{\rigmark}}
%\plainfootrulewidth 0.4pt
\newcommand{\be}{\begin{equation}}
\newcommand{\ee}{\end{equation}}
\newcommand{\bm}{\boldmath}
\newcommand{\ds}{\displaystyle}
\newcommand{\bea}{\begin{eqnarray}}
\newcommand{\eea}{\end{eqnarray}}
\newcommand{\ba}{\begin{array}}
\newcommand{\ea}{\end{array}}
\newcommand{\arcsinh}{\mathop{\rm arcsinh}\nolimits}
\newcommand{\arctanh}{\mathop{\rm arctanh}\nolimits}
\newcommand{\bc}{\begin{center}}
\newcommand{\ec}{\end{center}}

\thispagestyle{plain}

\label{sh}

%\lfoot[\fancyplain{\ \\[1mm] \thepage}{\ \\[1mm]\thepage}]{\fancyplain{}{}}

\begin{center} {\Large \bf
\begin{tabular}{c}
HERMITE POLYNOMIAL 
\\[-1mm]
REPRESENTATION OF THE SPIN STATES
\end{tabular}
 } \end{center}

\bigskip

\bigskip

\begin{center} {\bf
Dmitry B. Lemeshevskiy$^{1*}$ and Vladimir I. Man'ko$^{2}$  
}\end{center}

\medskip

\begin{center}
{\it
$^1$Moscow Institute of Physics and Technology (State University)\\
Institutskii per. 9, Dolgoprudnyi, Moscow Region 141700, Russia
\smallskip

$^2$P. N. Lebedev Physical Institute, Russian Academy of Sciences\\
Leninskii Prospect 53, Moscow 119991, Russia
}
\smallskip

$^*$Corresponding author e-mail:~~~d.lemeshevskiy@gmail.com,~~~manko@sci.lebedev.ru\\
\end{center}

\begin{abstract}\noindent
The invertable map of spin state density operator onto quasiprobability distribution of three continuous variables is constructed. The connection with two-mode electromagnetic field oscillators is discussed. The inversion formula for spin-density matrix is given in terms of Hermite polynomials. The connection with the spin-tomographic representation is examined. The problem of the magnetic moment rotation in the magnetic field is considered in the suggested representation of spin states.
\end{abstract}

\medskip


\section{Introduction}
\pst
Recently \cite{PRA2012} the Wigner function for spin states was studied by using Jordan-Schwinger \cite{Jordan, Schwinger} map. This map permits to construct the angular momentum theory in quantum mechanics due to its relation to two-dimensional oscillator energy-states \cite{BidLouck}. In \cite{PLA96, Ibort2009} the tomographic probability representation for quantum states for systems with continuous variables was suggested on base of Radon transform \cite{Rad} of standard Wigner function \cite{Wign32}. The spin tomograms for states with discrete variables was suggested in \cite{PLA97, JETP97}. The spin tomograms in this construction are the probability distributions with finite number of outcomes (spin-projections $m$).

There exist several different schemes where the spin states are identified with the probability distribution functions (see, e. g. \cite{Filippov2010}). The aim of this work is to construct new representation for spin states using the mentioned in \cite{PRA2012} approach based on two-dimensional oscillator connection with angular momentum theory. The paper is organised as follows. In Sec.2 we briefly review the star-product approach \cite{JETP57, JPA2002, PLA2005} to the description of the quantum states. The spin tomogram is reviewed as well. In Sec. 3 the definition of the new representation of the spin states is given. A short example of using the framework considered is examined in Sec. 4. Finally, in Sec. 5 the conclusions are given.

\section{Star-product approach to the spin-state description}
\pst
Let us consider the problem of the description of the quantum states of a system with the angular momentum $j$. Conventionally, the mutual system of eigenvectors of the operators $\hat J^2$ and $\hat J_z$ is chosen as the basis in the Hilbert spase of the states of the system:
\begin{equation}
\hat J^2|jm\rangle=j(j+1)|jm\rangle, \hat J_z|jm\rangle=m|jm\rangle,
\end{equation}
where $m$ is the spin projection on the quantization axis ($Oz$-axis) $(m=-j,-j+1,...,j).$

In the most general case the state of the system can be described using the density operator $\hat\rho^{(j)}$ which can be represented in terms of the matrix elements $\rho^{(j)}_{m_1m_2}=\langle jm_1|\hat\rho^{(j)}|jm_2\rangle$ in the following form:
\begin{equation}
\hat\rho^{(j)} = \sum_{m_1,m_2}\rho^{(j)}_{m_1m_2}|jm_1\rangle\langle jm_2|.
\end{equation}

On the other hand, one can find another representation of quantum states, where the states are identified with the functions of some set of the variables that we denote $\vec x$. This problem comes to discovering a pair of maps: the direct one from the density operator $\hat\rho$ to some function $f(\vec x)$ and the inverse one that acts vice-versa. A general rule that describes such pairs of maps can be formulated. This rule is formulated in terms of the dequantizer $\hat U(\vec x)$ and quantizer $\hat D(\vec x)$ operators and the whole task is to define these operators for a particular representation. The image of the operator after the direct transform is called the operator's symbol and for an operator $\hat A$ is denoted as $f_{\hat A}(\vec x)$. It is generally stated, that the direct transform has the following form:
\begin{equation}
\label{dequant}f_{\hat A}(\vec x)=\tr\left(\hat A\hat U(\vec x)\right).
\end{equation}

Besides the direct map the inverse one has also a general form. It implies integrating over the whole set of variables $\vec x$ that, in the case of a discrete variable in the set, should be treated as taking a sum. The inverse map has the following form:
\begin{equation}
\label{quant}\hat A = \int f_{\hat A}(\vec x)\hat D(\vec x)d\vec x.
\end{equation}

Once the formulas for the direct and inverse maps have been written down one should formulate the condition for the quantizer and dequantizer operators that will assure the consistency of the maps given by Eq. \eqref{dequant} and Eq. \eqref{quant}. This condition can also be formulated in the general form and can be treated as the orthogonality condition. But in some cases there is no use of the general formula and the consistency of a particular pair quantizer-dequantizer should be checked directly. In the case of spin tomogram and the representation described in this paper the direct check is appropriate.

The quantizer-dequantizer framework not only defines the form of the direct and inverse maps of the quantum mechanical operators but also describes the procedure of calculating the mean values of the different observables in the representation constructed. At this point we assume that the quantizer and dequantizer operators are defined and the consistency of the maps \eqref{dequant} and \eqref{quant} is proven. We denote the symbol of the density operator of the system $f_{\hat\rho}(\vec x)$. This function carries full information about the system's state. Let us consider a quantum mechanical operator $\hat A$ corresponding to an observable physical quantity. It is well known that in terms of the density operator $\hat\rho$ the mean value of the observable $< \hat A >$ is calculated according to the following rule:
\begin{equation}
\label{mean}< \hat A > = \tr(\hat\rho\hat A).
\end{equation}

Assuming that the density operator posess the integral expansion \eqref{quant} and changing the order of the integration and taking the trace operations in this expression is legal one can rewrite \eqref{mean} in the following form:
\begin{equation}
\label{mean1}< \hat A > = \int f_{\hat\rho}(\vec x)\tr(\hat A\hat D(\vec x))d\vec x.
\end{equation}

It is appropriate to introduce the notion of the dual symbol of an operator $\hat A$. By the definition, the dual symbol of the operator $\hat A$, denoted as $f_{\hat A}^d$, is calculated according to the following formula:
\begin{equation}
\label{dual}f_{\hat A}^d=\tr\left(\hat A\hat D(\vec x)\right).
\end{equation}

Comparing Eq. \eqref{mean1} and \eqref{dual} results into a concise formula for the calculating the mean values of the observables, that in some sense coincides with the clasical statistical definition of the mean value of a quantity as an integral of the product of this quantity and the probability distribution function:
\begin{equation}
<\hat A>=\int f_{\hat A}^d(\vec x)f_{\hat\rho}(\vec x)d\vec x.
\end{equation}
Though the analogy is obvious it is incorrect to assume that an arbitrary pair of the quantizer and dequantizer operators will result into a fair-probability distribution function $f_{\hat\rho}(\vec x)$ of a measureable quantum observable. If the symbol of the density operator does not satisfy some of the requirments for the probability distribution function it is sometimes called quasiprobability distribution. The examples of such quasiprobabilities are well-studied for the case of the systems with continuous variables. For example, the Wigner function \cite{Wign32}, Glauber-Sudarshan \cite{Glau63, Sudar63} P-function and Husimi Q-function \cite{Hus40} are referred to as quasiprobability distributions.

Another aspect of the star-product scheme is the posibility to find the relation between different representations of the same quantum-mechanical operators obtained in schemes with different pairs of dequantizer and quantizer. Let us consider two consistent schemes of star-product quantization. The first scheme uses the pair of dequantizer $\hat U^{(a)}(\vec x)$ and quantizer $\hat D^{(a)}(\vec x)$. The second scheme is described by the operators $\hat U^{(b)}(\vec y)$ and $\hat D^{(b)}(\vec y)$ correspondingly. The symbols of the physical operator $\hat A$ in the two schemes considered are $f_{\hat A}^{(a)}(\vec x)$ and $f_{\hat A}^{(b)}(\vec y)$. Our aim is to show how one of this symbols can be expressed in terms of the another. Applying the definition of the symbol $f_{\hat A}^{(a)}(\vec x)$ from the Eq. \eqref{dequant} and the inverse relation \eqref{quant} for the symbol $f_{\hat A}^{(b)}(\vec y)$ and assuming that changing the order of the integration and taking the trace operations is legal one can derive the following formula:
\begin{equation}
f_{\hat A}^{(a)}(\vec x)=\int\tr\left(\hat U^{(a)}(\vec x)\hat D^{(b)}(\vec y)\right)f_{\hat A}^{(b)}(\vec y)d\vec y=\int K_{ab}(\vec x,\vec y)f_{\hat A}^{(b)}(\vec y)d\vec y.
\end{equation}

We introduced the notation $K_{ab}(\vec x,\vec y)$ for the integral kernel of this map. The inverse relation can be obtained in a similar form with the integral kernel denoted $ K_{ba}(\vec y,\vec x)$ (one may note that difference of these two notations is in the subscripts order). To sum up, we write down the formulas for calculating both kernels:
\begin{equation}
\label{interkernels}K_{ab}(\vec x,\vec y)=\tr\left(\hat U^{(a)}(\vec x)\hat D^{(b)}(\vec y)\right), K_{ba}(\vec y,\vec x)=\tr\left(\hat U^{(b)}(\vec y)\hat D^{(a)}(\vec x)\right).
\end{equation}

The spin tomogram representation was formulated for the first time in the papers \cite{PLA97, JETP97}. The dequantizer operator has a simple form and, moreover, a clear physical meaning. The spin tomogram is denoted $\w(m,\alpha,\beta)$, where $\alpha$ and $\beta$ have the meaning of the Euler angles determining the point on the sphere $S^2$ and $m$ is the spin projection. The physical meaning of the tomogram is the probability distribution of the spin projection $m$ measured in the rotated reference frame. The Euler angles describe the refernece frame transformation or, in other words, the new direction of the quantization axis. One can describe the new direction of the quantization axis using a unit vector ${\mathbf n}(\alpha,\beta)=(\cos\alpha\sin\beta;\sin\alpha\sin\beta;\cos\beta)$ directed along it. Then the tomogram arguments are ${\mathbf n}(\alpha,\beta)$ and $m$. In these terms, the expressions for the spin tomogram and the dequantizer $\hat U(m, \alpha,\beta)$ have the following form:
\begin{equation}
\label{spintom}\w(m,\alpha,\beta)=\tr\left(\delta\left(m-\left({\mathbf n}(\alpha,\beta)\cdot\hat{\mathbf J}\right)\right)\hat\rho\right), \hat U(m, \alpha, \beta)=\delta(m-({\mathbf n}(\alpha,\beta)\cdot\hat{\mathbf J})).
\end{equation}
The inverse transform can be given in several different forms. But all these inverse transform expressions have an awkward form. We present one of them, which involves Clebsch-Gordan coefficients $\langle j_1m_1;j_2m_2|JM\rangle$ and Wigner $\mathcal D$-function $\mathcal D^{(j)}_{m_1m_2}(\alpha,\beta,\gamma)=e^{im_2\alpha}d^{(j)}_{m_1m_2}(\beta)e^{im_1\gamma}$ where
$$
d^{(j)}_{m_1m_2}(\beta)=\left[\frac{(j+m_1)!(j-m_2)!}{(j+m_2)!(j-m_1)!}\right]^{1/2}\left(\cos\frac\beta 2\right)^{m_1+m_2}\left(\sin\frac\beta 2\right)^{m_1-m_2}P_{j-m_1}^{(m_1-m_2, m_1+m_2)}(\cos\beta),
$$
$P_n^{(a,b)}(x)$ is the Jacobi polynomial and $j=0,1/2,1,3/2,...$.

Using these notations one can write down explicitly the matrix elements for the quantizer operator $D^{(j)}_{m_1m_2}(m,\alpha,\beta)$:
\begin{equation}
D^{(j)}_{m_1m_2}(m,\alpha,\beta)=(-1)^m\sum_{j'=0}^{2j}\sum_{m'=-j'}^{j'}(-1)^{m_1}\mathcal D_{0m'}^{(j')}(\alpha,\beta,\gamma)(2j'+1)\langle jm;j-m|j'0\rangle\langle jm_1;j-m_2|j'm'\rangle,
\end{equation}
and the inverse transform can be written down in the following form:
\begin{equation}
\rho^{(j)}_{m_1m_2}=\frac 1{4\pi}\int_0^{2\pi}d\alpha\int_0^{\pi}\sin\beta d\beta\sum_{m=-j}^j\w(m,\alpha,\beta)D^{(j)}_{m_1m_2}(m,\alpha,\beta).
\end{equation}

Once the formulas for an arbitrary $j$ were presented in a general form one should consider the simpliest particular case of the $j=1/2$. In this case both quantizer and dequantizer are square $2\times2$ matrices \cite{Filippov2009}. The dequantizer operator has the following representation:
\begin{equation}
\label{U-12}\hat U^{(1/2)}(m,\alpha,\beta)=\frac12\begin{pmatrix}1&&0\\0&&1\end{pmatrix}+m\begin{pmatrix}\cos\beta && -e^{i\alpha}\sin\beta\\  -e^{-i\alpha}\sin\beta && -\cos\beta\end{pmatrix}.
\end{equation}
On the other hand, the quantizer operator can be obtained from the expression above by multiplying second term in the formula \eqref{U-12} by 3:
\begin{equation}
\label{D-12}\hat D^{(1/2)}(m,\alpha,\beta)=\frac12\begin{pmatrix}1 && 0 \\  0 && 1\end{pmatrix}+3m\begin{pmatrix}\cos\beta && -e^{i\alpha}\sin\beta\\  -e^{-i\alpha}\sin\beta && -\cos\beta\end{pmatrix}.
\end{equation}

\section{The Hermite polynomial representation of the spin states}
\pst
In this section we introduce a new representation of the quantum spin states by defining appropriate pair quantizer-dequantizer. All the results presented in this section have been achieved due to the following initial idea. The well-known Jordan-Schwinger map allows to treat the spin states of the system with an arbitrary total angular momentum as corresponding Fock states of a two-dimensial oscillator. On the other hand, the oscillator is a system studied in details in various representations of the quantum mechanics. Among these representations one can find both optical and symplectic tomography approaches. The tomographic approach is an example of the start-product quantization scheme that was briefly described in Sec. 2 of this paper. Combining these two maps allows to construct a new representation of the spin states that we will call the Hermite polynomial representation or simply the H-representation because both the quantizer and dequantizer operators contain the Hermite polynomials in their definitions.

We start the description of the H-representation from the definition of the following set of states, identified with three continuous variables $x,y$ (which can take any real value) and angle $\theta$ (which takes values from the domain $0$ to $2\pi$):
\begin{equation}
\label{state}|x,y,\theta\rangle^{(j)}=\frac{1}{2^{j}\sqrt{\pi}}\exp\left(-\frac{x^2+y^2}2\right)\sum_{m=-j}^je^{im\theta}\frac{H_{j+m}(x)H_{j-m}(y)}{\sqrt{(j-m)!(j+m)!}}|jm\rangle,
\end{equation}
where $H_k(x)$ is the Hermite polynomial, defined as follows:
\begin{equation}
H_k(x)=(-1)^ne^{x^2}\frac{d}{dx^k}e^{-x^2}.
\end{equation}

In terms of the introduced states we define the dequantizer operator $\hat U^{(j)}(x,y,\theta)$ in the following way:
\begin{equation}
\label{dequantizer}\hat U^{(j)}(x,y,\theta)=|x,y,\theta\rangle^{(j)}\langle x,y,\theta|^{(j)}.
\end{equation}

At this point we are ready to introduce the H-representation of the spin states. Following the procedure described by the Eq. \eqref{dequant} we define the function $\w (x,y,\theta)$ which we will call the H-distribution:
\begin{equation}
\label{direct}\w (x,y,\theta)=\tr\left(\hat\rho^{(j)}\hat U^{(j)}(x,y,\theta)\right).
\end{equation}

For the pure states, that can be described using wave-function or state-vector $|\psi\rangle$ this formula reduces to a simplier one:
\begin{equation}
\label{direct1}\w (x,y,\theta)=\left|\langle\psi|x,y,\theta\rangle\right|^2.
\end{equation}

Once we have introduced the H-representation we need to show that this representation of the spin states is informationally complete. In other words, we have to define the quantizer $\hat D^{(j)}(x,y,\theta)$ and show that transform \eqref{direct} is invertable, so the density operator $\hat\rho^{(j)}$ can be reconstructed from the known H-distribution $\w(x,y,\theta)$.

We define the quantizer $\hat D^{(j)}(x,y,\theta)$ by writing down it's matrix elements:
\begin{equation}
\label{quantizer}D_{mm'}(x,y,\theta)=\frac{e^{i(m-m')\theta}}{2^{2j+1}\pi}\frac{\sqrt{(j+m)!(j+m')!(j-m)!(j-m')!}}{(2j+m+m')!(2j-m-m')!}H_{2j+m+m'}(x)H_{2j-m-m'}(y).
\end{equation}

Then the inverse transform is represented by the following integral:
\begin{equation}
\label{reverse}\hat\rho^{(j)}=\int_0^{2\pi}d\theta\int_{-\infty}^{\infty}dx\int_{-\infty}^{\infty}dy\w (x,y,\theta)\hat D^{(j)}(x,y,\theta).
\end{equation}

As it has been already said in the Sec. 2 in the case of the introduced H-representation it is appropriate to check directly the consistency of the scheme defined by Eq. \eqref{direct} and Eq. \eqref{reverse}. One should start with another form of the expression \eqref{direct} where the density matrix $\hat\rho^{(j)}$ is expressed in the terms of the matrix elements $\rho^{(j)}_{m_1m_2}$:
\begin{equation}
\w (x,y,\theta)=\frac{e^{-(x^2+y^2)}}{2^{2j}\pi}\sum_{m_1,m_2=-j}^j\rho^{(j)}_{m_1m_2}e^{-i(m_1-m_2)\theta}\frac{H_{j+m_1}(x)H_{j+m_2}(x)H_{j-m_1}(y)H_{j-m_2}(y)}{\sqrt{(j+m_1)!(j+m_2)!(j-m_1)!(j-m_2)!}}.
\end{equation}

Comparing this expression with the quantization procedure given by Eq. \eqref{reverse} one conclude that the integral of the following form has to be evaluated:
$$
\int_{-\infty}^{\infty}e^{-z^2}H_p(z)H_n(z)H_k(z)dz.
$$
The result of calculations can be expressed as follows \cite{Busbridge}. If the following conditions are satisfied (here $s$ is a nonnegative integer number):
$$
p+n+k=2s, p\leqslant s, n\leqslant s, k\leqslant s, 
$$
then the value of the integral is expressed in terms of $p,n,k,s$ in the following form:
\begin{equation}
\label{hermite}\int_{-\infty}^{\infty}e^{-z^2}H_p(z)H_n(z)H_k(z)dz = \frac{\sqrt\pi2^sp!n!k!}{(s-p)!(s-n)!(s-k)!}.
\end{equation}
Otherwise, the integral is equal to $0$.

The proof continues as follows: one should take the matrix element $\langle jm|\bullet|jm'\rangle$ of the expression \eqref{reverse}, and perform integrating over the $d\theta$ that results into a factor proportional to $\delta_{m-m',m_1-m_2}$. Next, integral over the $dx$ doesn't vanish just if the condition $m_1\geqslant m$ is satisfied. On the other hand, integral over the $dy$ doesn't vanish if $m_1\leqslant m$. The combination of these three conditions can be treated as the acting of a delta-symbol $\delta_{m,m_1}\delta_{m',m_2}$. This statement finishes the proof.

Our next point is to describe some properies of the H-distribution. First, for any value of the variables $x,y,\theta$ the H-distribution $\w(x,y,\theta)$ posesses only real and nonnegative values. This statement is a consequence of the expression \eqref{dequantizer} that defines the dequantizer operator $\hat U^{(j)}(x,y,\theta)$.

Second, the H-distribution $\w(x,y,\theta)$ satisfies the normalization condition in the following form (this expression is valid for any value of the variable $\theta$):
\begin{equation}
\int_{-\infty}^{\infty}dx\int_{-\infty}^{\infty}dy\w (x,y,\theta)=1.
\end{equation}

To prove the fact that H-distribution is normalized in this sense one should start from the considering the following relation of completness of the system of states $|x,y,\theta\rangle$:
\begin{equation}
\int_{-\infty}^{\infty}dx\int_{-\infty}^{\infty}dy|x,y,\theta\rangle\langle x,y,\theta|=\sum_{m=-j}^{j}|jm\rangle\langle jm|=\hat 1,
\end{equation}
where the integral is evaluated in accordance with the well-known orthogonality relation for the Hermite polynomials $H_n(z)$:
$$
\int_{-\infty}^{\infty}e^{-z^2}H_n(z)H_k(z)dz=2^nn!\sqrt\pi\delta_{nk}.
$$

Nonnegativity and normalization of the H-distribution \eqref{direct} gives a posibility to interpret it as a fair probability distribution function. To make this interpretation physically clear one needs to say probability of what events describes this function. At the moment this problem needs extra clarification. In view of this we will interpret the H-distribution as one of spin quasiprobability distributions, analogues to the case of Husimi function used to describe the systems with continuous variables.

Moreover, in case the H-distribution represents a fair probability distribution one can interpret it as the conditional probability distribution \cite{AIP}, i.e. using standart notation we have:
\begin{equation}
\w(x,y,\theta)\equiv\w(x,y|\theta).
\end{equation}

It means that one can introduce the joint probability distribution $P(x,y,\theta)$ applying Bayesian formula:
\begin{equation}
P(x,y,\theta)=\mathcal P(\theta)\w(x,y|\theta)
\end{equation}
where $\mathcal P(\theta)$ is an arbitrary distribution on the circle $0\leqslant\theta\leqslant 2\pi$. The physical meaning of this interpretation needs extra clarification.

\section{Example: magnetic moment rotation in the magnetic field}
\pst
In this section we give a simple example that shows how a particular physical problem can be described in the terms of the H-distribution. We consider an electron with the spin $1/2$ in the constant magnetic field. At the initial moment of time $t=0$ the electron's spin is oriented in the direction of the $Oz$ axis. Magnetic field $\mathbf{B}$ is oriented in the direction of the $Ox$ axis. Our aim is to find the dependence on time of the H-distribution that describes this physical situation. In order to do this we will use the solution of the Schr\"odinger equation for the evolution operator ${\mathcal {\hat U}}(t)$. The Schr\"odinger equation for the evolution operator reads:
\begin{equation}
 i\hbar\frac{\partial{\mathcal {\hat U}}(t)}{\partial t}=\hat H{\mathcal {\hat U}}(t),
\end{equation}
where the Hamiltonian of the system is expressed in the following form:
\begin{equation}
\hat H=-(\hat{\mathbf{\mu}},\mathbf B)=\frac{\hbar\omega_c}2\begin{pmatrix}0 && 1\\1 && 0\end{pmatrix},
\end{equation}
and the following notation for the frequency $\omega_c$ was used:
\begin{equation}
\omega_c=\frac{|e|B}{m_ec}.
\end{equation}

We write down explicitly the solution of the Schr\"odinger equation for the evolution operator. It reads:
\begin{equation}
{\mathcal {\hat U}}(t)=\begin{pmatrix}\cos\frac{\omega_c t}2&&-i\sin\frac{\omega_c t}2\\-i\sin\frac{\omega_c t}2&&\cos\frac{\omega_c t}2\end{pmatrix}.
\end{equation}

Using the expression for the evolution operator one can find the dependence on time of the spin state vector and the density matrix that corresponds to this state:
\begin{equation}
|\psi(t)\rangle={\mathcal {\hat U}}(t)\begin{pmatrix}1\\0\end{pmatrix}=\begin{pmatrix}\cos\frac{\omega_c t}2\\-i\sin\frac{\omega_c t}2\end{pmatrix},\hat\rho(t)=\begin{pmatrix}\cos^2\frac{\omega_c t}2&&i\sin\frac{\omega_c t}2\cos\frac{\omega_c t}2\\-i\sin\frac{\omega_c t}2\cos\frac{\omega_c t}2&&\sin^2\frac{\omega_c t}2\end{pmatrix}.
\end{equation}

Now we need to write down explicitly the formulas for quantizer and dequantizer for the case of the spin $1/2$. Using the definitions given by Eq. \eqref{state}, \eqref{quantizer} of quantizer and dequantizer and \eqref{hermite} for Hermite polynomials one obtain:
\begin{gather}
\label{hdequant}|x,y,\theta\rangle=\sqrt{\frac{2}{\pi}}\exp\left(-\frac{x^2+y^2}2\right)\begin{pmatrix}xe^{i\theta/2}\\ye^{-i\theta/2}\end{pmatrix},\hat U^{(1/2)}(x,y,\theta)=\frac{2}{\pi}e^{-(x^2+y^2)}\begin{pmatrix}x^2&&xye^{i\theta}\\xye^{-i\theta}&&y^2\end{pmatrix}\\
\label{hquant}\hat D(x,y,\theta)=\frac{1}{4\pi}\begin{pmatrix}2x^2-1 && 4xye^{i\theta} \\  4xye^{-i\theta} && 2y^2-1\end{pmatrix}.
\end{gather} 

It is worthy to note that in this particular case for the system with total angular momentum $j=1/2$ the expression for the states $|x,y,\theta\rangle^{(1/2)}$ can be reduced to a form that is very similar to the expression for the coherent spin states \cite{Radcliffe}. In order to do this one should consider introducing polar coordinates $(r,\varphi)$ in the $(x,y)$ plane. The two angles $\varphi$ and $\theta$ would identify the coherent spin state. The $|x,y,\theta\rangle^{(1/2)}$ state differs from the coherent spin state in the normalization factor. As it can be easily seen, the states $|x,y,\theta\rangle^{(1/2)}$ are not normalized and in the case of introducing polar coordinates the normalization factor is a function of the $r$ variable only. Thus, the analogy between the H-representation for the spin states and Husimi representation for the states of systems with continuous variables is more clear.

For the systems with total angular momentum equal to $1/2$ one can demonstrate the connection of the symbols of the same operators in the H-representation and the spin-tomogram representation. We denote the spin tomogram distribution function for the spin $1/2$, defined in the Eq. \eqref{spintom} as $\w^{(1/2)}_{(T)}(m,\alpha,\beta)$ and the H-distribution as $\w^{(1/2)}_{(H)}(x,y,\theta)$. The explicit formulas for the kernels $K_{TH}^{(1/2)}(m,\alpha,\beta,x,y,\theta)$ and $K_{HT}^{(1/2)}(x,y,\theta,m,\alpha,\beta)$ obtained from the definitions \eqref{interkernels} and explicit expressions for the dequantizers \eqref{U-12}, \eqref{hdequant} and quantizers \eqref{D-12}, \eqref{hquant} for both schemes read:
\begin{gather}
K_{TH}^{(1/2)}(m,\alpha,\beta,x,y,\theta)=\frac{1}{4\pi}\left(x^2+y^2-1+2m\left[(x^2-y^2)\cos\beta-4xy\sin\beta\cos(\alpha-\theta)\right]\right),\\
K_{HT}^{(1/2)}(x,y,\theta.m,\alpha,\beta)=\frac{1}{\pi}e^{-(x^2+y^2)}\left(x^2+y^2+6m\left[(x^2-y^2)\cos\beta-2xy\sin\beta\cos(\alpha-\theta)\right]\right).
\end{gather}

These two integral kernels can be used to find one of the functions $\w^{(1/2)}_{(T)}(m,\alpha,\beta)$ or $\w^{(1/2)}_{(H)}(x,y,\theta)$ if the another is known. The explicit formulas for such calculations are given by the following two equations:
\begin{gather}
\w^{(1/2)}_{(T)}(m,\alpha,\beta)=\int_0^{2\pi}d\theta\int_{-\infty}^{\infty}dx\int_{-\infty}^{\infty}dyK_{TH}^{(1/2)}(m,\alpha,\beta,x,y,\theta)\w^{(1/2)}_{(H)}(x,y,\theta),\\
\w^{(1/2)}_{(H)}(x,y,\theta)=\frac 1{4\pi}\int_0^{2\pi}d\alpha\int_0^{\pi}\sin\beta d\beta\sum_{m=-j}^jK_{HT}^{(1/2)}(x,y,\theta.m,\alpha,\beta)\w^{(1/2)}_{(T)}(m,\alpha,\beta).
\end{gather}

These relations mean that quasiprobability H-distribution can be mapped onto spin-tomographic fair-probability distribution. It is worthy to mention that other quasiprobabilities of spin states were also shown to be mapped explicitly onto spin-tomogram in \cite{Scully}.

Using the explicit formula for the $|x,y,\theta\rangle^{(1/2)}$ state and transform \eqref{direct1} or alternatively the expression for the dequantizer $\hat U^{(1/2)}(x,y,\theta)$ and the transform \eqref{direct} one can find the expression for the H-distribution for the problem considered:
\begin{multline}
\label{tom}\w(x,y,\theta,t)=\frac{2}{\pi}e^{-(x^2+y^2)}\left(x^2\cos^2\frac{\omega_c t}2+2xy\sin\frac{\omega_c t}2\cos\frac{\omega_c t}2\sin\theta+y^2\sin^2\frac{\omega_c t}2\right)=\\
=\frac{1}{\pi}e^{-(x^2+y^2)}\left(x^2+y^2+ (x^2-y^2)\cos\omega_c t+2xy\sin\omega_c t\sin\theta\right).
\end{multline}

To calculate the dependence on time of the mean values of the spin projections $<\hat S_x>,<\hat S_y>, <\hat S_z>$ we start from the explicit formulas for the dual symbols of these operators. According to the definition \eqref{dual} of the dual symbol and the expression \eqref{hquant} for the dequantizer one can find the following expressions for the dual symbols of the spin projections:
\begin{equation}
f_x(x,y,\theta)=\frac{1}{\pi}xy\cos\theta,\qquad f_y(x,y,\theta)=-\frac{1}{\pi}xy\sin\theta, \qquad f_z(x,y,\theta)=\frac{1}{4\pi}(x^2-y^2).
\end{equation}

Now one can integrate these functions with the H-distribution function \eqref{tom} to obtain the dependence on time of the mean values. After some algebra one obtains the result:
\begin{equation}
<\hat S_x> = 0, \qquad <\hat S_y> = -\frac12\sin\omega_c t, \qquad <\hat S_z> = \frac12\cos\omega_c t.
\end{equation}

These results fully coincide with the well-known solution and represent the rotation of the mean value of the electron's spin around the direction of the magnetic field.
\section{Summary}
\pst
We point out the main results of our work. New star-product scheme based on Hermite polynomial properties to describe spin states was introduced. Quantizer and dequantizer containing the dependence on coordinates of an "artificial oscillator" were found in explicit form. The function which is symbol of the spin density operator was shown to be nonnegative and normalised with respect to the two-mode oscillator coordinates. In this sense this function has the properties analogues to tomographic probability distribution describing states of quantum systems. We call the representation of spin states suggested in this paper as Hermite polynomial representation, or H-representation. The problem to interpret the function as fair probability distribution will be studied in future paper as well as posibility to apply the developed scheme to describe the multilevel system states.

\section*{Acknowledgments}
\pst This study was partially supported by the Russian Foundation for Basic Research under Projects Nos. 10-02-00312 and 11-02-00456.

\end{document}